\documentclass[12pt,a4paper]{article}

\usepackage{enumitem}
\usepackage{pifont}
\usepackage{pdfpages}
\usepackage{tcolorbox}
\usepackage{url}
\usepackage{amsmath,amssymb}
\usepackage{authblk}
\usepackage{pslatex}
\usepackage[top=2cm, bottom=2cm, left=2cm, right=2cm]{geometry}

\begin{document}

\title{\bf Predictability: Can the turning point and end of an expanding epidemic be precisely forecast while the epidemic is still spreading?}

\author[a,b]{Mario Castro}
\author[a,c]{Sa\'ul Ares}
\author[a,d,e,f]{Jos\'e A. Cuesta}
\author[a,c]{Susanna Manrubia}

\affil[a]{\small Grupo Interdisciplinar de Sistemas Complejos (GISC), Madrid, Spain}
\affil[b]{Instituto de Investigaci\'on Tecnol\'ogica (IIT), Universidad Pontificia Comillas, Madrid, Spain}
\affil[c]{Dept.~Biolog\'{\i}a de Sistemas, Centro Nacional de Biotecnolog\'ia (CSIC). c/ Darwin 3, 28049 Madrid, Spain}
\affil[d]{Dept.~Matem\'aticas, Universidad Carlos III de Madrid, Avenida de la Universidad 30, 28911 Leganes, Spain}
\affil[e]{Instituto de Biocomputaci\'on y F\'isica de Sistemas Complejos (BIFI), c/ Mariano Esquillor, Campus R\'io Ebro, Universidad de Zaragoza, 50018 Zaragoza, Spain}
\affil[f]{UC3M-Santander Big Data Institute (IBiDat), c/ Madrid 135, 28903 Getafe, Spain}

\date{}

\maketitle

\abstract{No, they can't. Epidemic spread is characterized by exponentially growing dyna\-mics, which are intrinsically unpredictable. The time at which the growth in the number of infected individuals halts and starts decreasing cannot be calculated with certainty before the turning point is actually attained; neither can the end of the epidemic after the turning point. An SIR model with confinement (SCIR) illustrates how lockdown measures inhibit infection spread only above a threshold that we calculate. The existence of that threshold has major effects in predictability: A Bayesian fit to the COVID-19 pandemic in Spain shows that a slow-down in the number of newly infected individuals during the expansion phase allows to infer neither the precise position of the maximum nor whether the measures taken will bring the propagation to the inhibition regime. There is a short horizon for reliable prediction, followed by a dispersion of the possible trajectories that grows extremely fast.  The impossibility to predict in the mid-term is not due to wrong or incomplete data, since it persists in error-free, synthetically produced data sets, and does not necessarily improve by using larger data sets. Our study warns against precise forecasts of the evolution of epidemics based on mean-field, effective or phenomenological models, and supports that only probabilities of different outcomes can be confidently given.}



\section{Introduction}

In 1972, Edward Norton Lorenz delivered a legendary-by-now talk entitled {\it Predictability: Does the Flap of a Butterfly’s Wings in Brazil Set Off a Tornado in Texas?} \cite{lorenz:1972}. Lorenz had stumbled onto chaos and uncovered its major consequence for weather prediction. By means of a simple model \cite{lorenz:1963}, he had shown that one can never be certain of whether, one week from now, we will have a sunny or a rainy day. Half a century later, we are used to listen to the weather forecast in terms of percentages, probability of rain, intervals for temperature and wind speed, and so on. Just fuzzy information, but usually sufficient to make up our minds on what to do next weekend. The key point is that, as far as weather is concerned, we accept that we are bound to cope with uncertainty. The mechanism behind that uncertainty has to do with the exponential amplification of small initial differences prototypical of chaotic systems. It turns out that other systems with exponentially growing variables also display an analogous behavior: they are sensitive to small variations in parameters and amplify small differences, potentially leading to quantitatively and qualitatively different outcomes. Though their dynamics are not chaotic, this is the case of epidemics. 

The world-wide on-going COVID-19 pandemic 
is triggering multiple attempts at modeling the progression and immediate future of epidemic spread \cite{estrada:2020,flaxman2020estimating,ivorra:2020,maier:2020,perez-garcia:2020,wong:2020}. Many of the formal approaches used are based on simple, mean-field compartmental models \cite{kermack:1927,hethcote:2000} with a different number of classes for the individuals in a population: susceptible ($S$), infected asymptomatic ($E$), infected symptomatic ($I$), recovered ($R$), dead ($D$) and several other possible intermediate stages such as quarantined, hospitalized or at the intensive care unit (ICU). Beyond their clear interpretation and ease of use, a main motivation to apply such models relies in trying to estimate the forthcoming stages of the epidemic, and in quantifying the effects of non-pharmaceutical measures towards ``flattening the curve'', ``reaching the peak'', estimating the total number of infected people when the epidemic ends, or controlling the number of ICUs required at a time from now. 

SIR-like models, therefore, are not only employed to drive intuitions and expectations, but are also applied to derive quantitative predictions. Further, the family of compartmental models lies at the basis of more sophisticated attempts to numerically describe the current spread of SARS-CoV-2 and the effect of contention measures \cite{prem:2020,wong:2020}, where they are mostly used at the local level \cite{gatto:2020,arenas:2020,li:2020_Sci}.

Here we consider a variant of an SIR model with reversible confinement of susceptible individuals that we call SCIR. But before we get into further details we would like to clarify what this paper is not: it is \emph{not} another paper with a simple model aiming to predict the evolution of the epidemic. On the contrary, we intend to show that predicting with models like this one is severely limited by strong instabilities with respect to parameter values. The reason we work with such a minimal model is that we can obtain analytical expressions for the dynamics under sensible approximations and make the point more clear. Also, it allows us to derive simple facts about the epidemics, such as the existence of a threshold that separates ``mild'' confinement measures causing mitigation from stronger measures leading to the inhibition of infection propagation.

The parameters of the model can be estimated within a relatively narrow range using data available from the COVID-19 pandemic. Yet, unavoidable uncertainties in those parameters, which determine the time at which growth is halted or the overall duration of the pandemic, propagate to the predicted trajectories, preventing reliable prediction of the intermediate and late stages of epidemic spread. This is the main message of this article, because it transcends the model of choice.

Attempts similar to the one performed here are common these days, and often strong predictions regarding the number of casualties, the position of the peak, or the duration of the epidemic are drawn. Our model does an excellent job in reproducing past data but, instead of taking most likely parameter values (or empirically evaluated values) to draw a prediction, we estimate compatible ranges of variations in the parameters. Especially when the process is close to the mitigation-inhibition threshold, predictions of the next few days become extremely sensitive to changes in the parameters, and to the addition of subsequent empirical data. Altogether, it turns out that quantitative predictions made in any similar framework are not reliable if not accompanied by their likelihood. The main conclusion we reach is that the deterministic nature of SIR-like models is misleading if aimed at describing the actual course of any pandemic: prediction of the past is achieved through suitable fitting of data, and different functions may work, but prediction of the future in the mid-term cannot be trusted. 

\section*{SCIR: an SIR model with confinement}

\begin{figure}[!ht]
    \centering
    \includegraphics[width=0.45\textwidth]{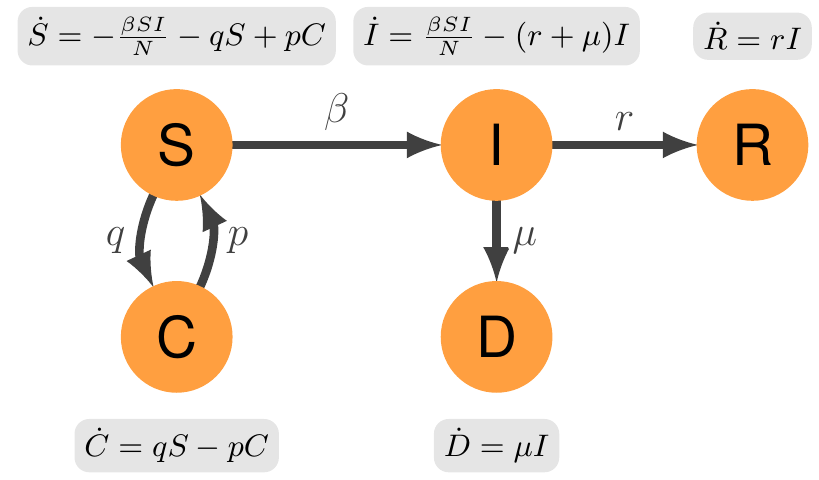}
\caption{\small Diagram of the epidemic model along with the equations ruling the dynamics. \emph{Susceptible} individuals ($S$) can enter and exit \emph{confinement} ($C$), or become \emph{infected} ($I$). Infected individuals can \emph{recover} ($R$) or \emph{die} ($D$). $N$ is the total population. Rates for each process are displayed in the figure, $q$ depends on specific measures restricting mobility and contacts, while $p$ stands for individuals that leave the confinement measures (e.g. people working at essential jobs like food supply, health care, policing) as well as for defection. We fit $I$ to data on officially diagnosed cases, which are automatically quarantined: the underlying assumption is that the real, mostly undetected, number of infections is proportional to the diagnosed cases.}
    \label{fig:diagram}
\end{figure}

The SCIR model includes the usual states of an SIR model plus a new class $C$ for individuals sent to confinement that are susceptible but not infected, see Fig.~\ref{fig:diagram}.

In a sufficiently large population, the number of infected individuals at the initial stages of the infection is well below the population size. Under certain conditions, it may stay small in comparison to the number of susceptible individuals remaining. This seems to be the case in July 2020 for most countries in the time elapsed since COVID-19 started to spread \cite{pollan2020prevalence,stringhini2020seroprevalence}.
If we assume that $I(t)/N\ll 1$, then we can neglect the nonlinear term in the equation for the number of susceptible individuals and solve the model analytically (see Materials and Methods and Supplementary Information, Section~A). Within this approximation, the number of infected individuals at time $t$ is given by
\begin{equation}
I(t)=I_0e^{[R^*_0(t)-1](r+\mu)t},
    \label{eq:linearSQX}
\end{equation}
where
\begin{equation}
R^*_0(t)=\frac{R_0}{q+p}\left[p+q\frac{1-e^{-(q+p)t}}{(q+p)t}\right],
\qquad R_0\equiv\frac{\beta}{r+\mu},
\label{eq:R0}
\end{equation}
is the effective basic reproduction number modulated by the confinement---$R_0$ being its value at the beginning of the epidemic. All the behavior of the epidemic is enclosed in this magnitude. At its initial stages $I(t)\sim\exp\big[(R_0-1)(r+\mu)t\big]$, so the epidemic spreads when $R_0>1$ (as is the case of COVID-19), and the larger $R_0$ the faster it does. When confinement sets in, $R^*_0(t)$ gets tamed, eventually dropping to the value $R^*_0(\infty)=R_0p/(q+p)$. An important epidemiological message follows from this simple fact: only if the confinement is strong enough ($p$ and $q$ are sufficiently different so that $R^*_0(\infty)<1$) can the epidemic be controlled; otherwise it spreads until eventually decaying due to the standard SIR mechanism---the exhaustion of susceptible individuals. 

Another interesting result follows from this simple model. Beyond the threshold for inhibition of infection spread, \eqref{eq:R0} captures the transient sub-exponential growth in the number of infected individuals. As mentioned above, if global confinement is suppressed ($q=p=0$) $I(t)$ grows exponentially at a rate $(R_0-1)(r+\mu)$. As confinement is turned on, $I(t)$ displays a systematic bending that, for long enough time, will lead to a second exponential regime characterized by the rate $[R^*_0(\infty)-1](r+\mu)$---which can be positive or negative depending on the confinement parameters. The bending of the curve is observed in both scenarios, so it cannot be taken as a sign that the epidemic will be eventually controlled.

\section*{Fitting COVID-19 data for Spain}

In order to illustrate the suitability of our model to reproduce available data, we have used official daily records reported by the Spanish Ministry of Health for all Spanish Autonomous regions since February 28th.  Strict lockdown permitting only essential trips outside the home was applied on March 14th. However, school and university closure took place on March 11th, so we take this date as the starting point of the confinement. The measure was extended on March 30th to the closure for two weeks of all businesses and companies not providing key services. 
Between these dates the data span two different regimes: unconstrained propagation of the epidemic, with $q=p=0$, and a lockdown phase with effective parameters for the transition to the confined state. Since separated data for the number of recovered and deaths was  unreliable, we have merged these two compartments and jointly fitted $r+\mu$, which become a single variable in practice.

We use a Bayesian approach to fit the data, assuming that the numbers of infected and recovered + dead are log-normally distributed with unknown variance and mean given by the expression for $I(t)$ obtained from the model (see Methods for details). At the very early stages of the epidemic (before any recovery or death event), the total number of confirmed cases grows as $e^{\beta t}$ independently of the chosen model (SIR, SEIR, etc.). Analyzing this initial growth for every country in the world, it appears that $\beta <1$ everywhere (doubling times larger than 1 day are reported in all cases~\cite{WorldInData}). Thus, we use informative priors for $\beta$ and $r+\mu$ (uniform distributions from $0$ to $1$ days$^{-1}$) and vague priors for the rates $q$ and $p$ (uniform distributions from $0$ to $5$ days$^{-1}$). Also, we use non-informative priors for the variances. 

\begin{figure}[!ht]
    \centering
    \includegraphics[width=.48\textwidth]{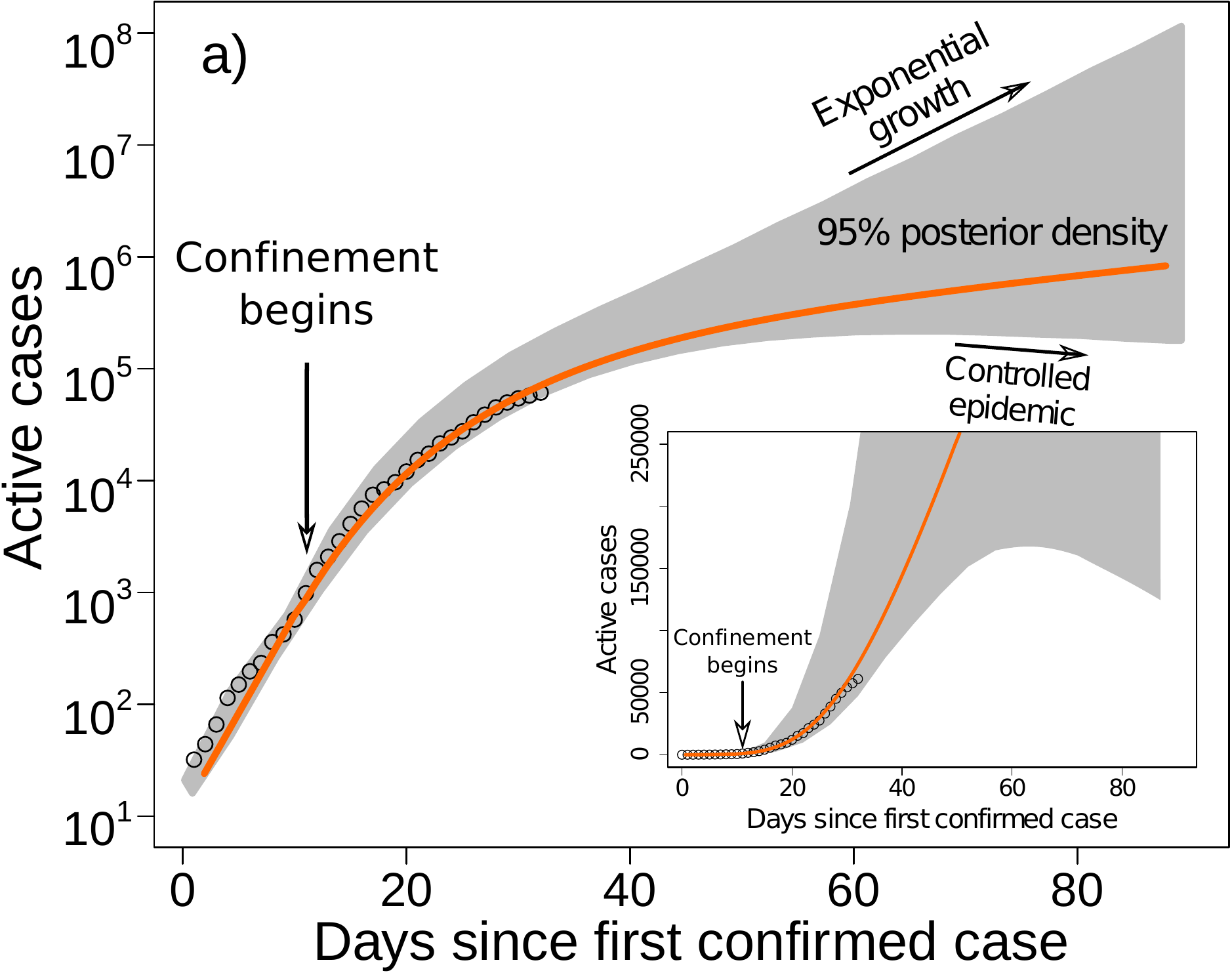}
    \includegraphics[width=.48\textwidth]{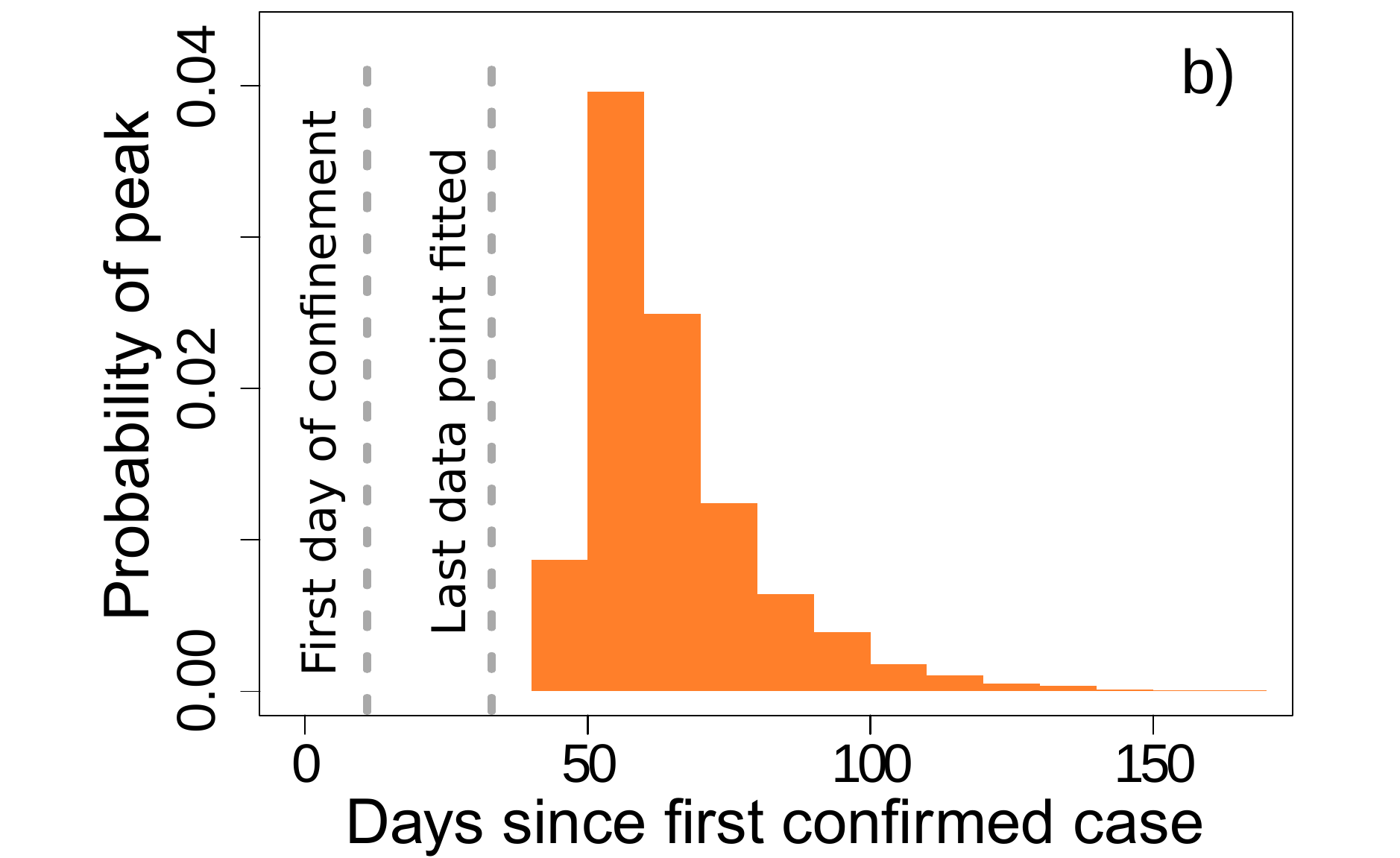}
    \caption{\small Fit to data obtained in real time for the daily number of active cases in Spain (from March 1st to March 29th) and peak forecast. (a) Despite the reasonable agreement between model and empirical observations in the growing phase, opposite predictions for the future number of active cases can be derived.
    The solid line represents the expression for $I(t)$ using the median parameters for each posterior in Fig.~S1 in SI. The vertical arrow denotes March 11th, the day when schools and universities closed.
    The shaded area represents the 95\% predictive posterior interval: its increasing width implies that predictability decays exponentially fast.
    Inset: same data and curves with linear vertical scale. {Figs.~S6 and S7 show how this fit and its posteriors evolve as an increasing number of days is included in the fit. An animation is included as Supplementary Movie.}
    (b) Posterior distribution of the time to reach the peak of the epidemic, \textit{conditioned} to actually having a peak (which occurs with probability 0.26). The vertical dashed lines stand for the days when the confinement began and for the date of the last data point used in the fit.
    }
   \label{fig:active-real}
\end{figure}

\subsection*{Uncertainty on peak occurrence fitting pre-peak data}

The results of fitting real-time data until March 29th are summarized in Figure~\ref{fig:active-real}. Figure~\ref{fig:active-real}(a) illustrates the fit to our analytical solution for the aggregated data of all Spanish Autonomous regions, representing country-level progression. Symbols are reported data, and the solid line represents the median of the distribution. Interestingly, quantiles 2.5\% and 97.5\% provide almost opposite conclusions: either the epidemic curve ``flattens'' or it keeps growing exponentially, albeit at a different rate. This is a consequence of the inherent variability of the fitted parameters ---as summarized by their posterior distributions (see Fig.~S1 in the SI)---and the exponential character of the epidemic. Similar conclusions can be drawn by inspection of the number of new deaths and recovered cases, $\Delta D+\Delta R$ (Fig.~S2 in SI). For completeness, we have also considered less realistic assumptions for the prior distributions and show that they lead to less consistent predictions. The obtained fits and posterior distributions are represented in Figs.~S3 and S4 of the Supplementary Information, respectively.

The systematic bending of the curve (see Fig.~\ref{fig:active-real}(a)), due to confinement in the framework of our model, does not guarantee that the epidemic is under control---hence, this information alone can be misleading in interpreting the effects of the measures applied. To emphasize this conclusion, we compute the posterior distribution of the time when the peak of the epidemic occurs. Analytically,
\begin{equation}
    t_{\max}= \frac{1}{p+q}\log \left(\frac{\beta q}{(r+\mu)(p+q)-\beta p}\right),
    \label{eq:tmax}
\end{equation}
which of course is only meaningful when the epidemic gets eventually controlled by the confinement measures (i.e., if $(r+\mu)(p+q)>\beta p$). With parameter values inferred from Fig.~\ref{fig:active-real}(a), confinement measures succeed at inhibiting the epidemic---which is the effect sought---only in $26\%$ of cases, while in $74\%$ of cases they fail at inhibiting its expansion and only slow it down. Figure~\ref{fig:active-real}(b) displays the distribution of the day in which the epidemic reaches the maximum, conditional on it actually occurring.

We have also fitted the model to each Spanish Autonomous region and have obtained analogous overall conclusions. Fits to reported data can be seen in Fig.~S5; all of them illustrate the goodness of our simple SCIR model at fitting past data. Posterior distributions yield comparable parameter values, within their intervals of definition, though there is significant variability in the percent of trajectories compatible with inhibition of propagation in each Autonomous region. First, the closer a region is to the inhibition threshold, the larger the dispersion of the forecast. In such Autonomous regions the epidemic started sooner than in regions that, by the end of March, were clearly in an expanding phase. Secondly, there are multiple regions with a vanishing probability of inhibition under the conditions of the first lockdown applied.

\subsection*{Uncertainty of epidemic end including post-peak data}

\begin{figure}[!ht]
    \centering
		\includegraphics[width=0.45\textwidth,clip=]{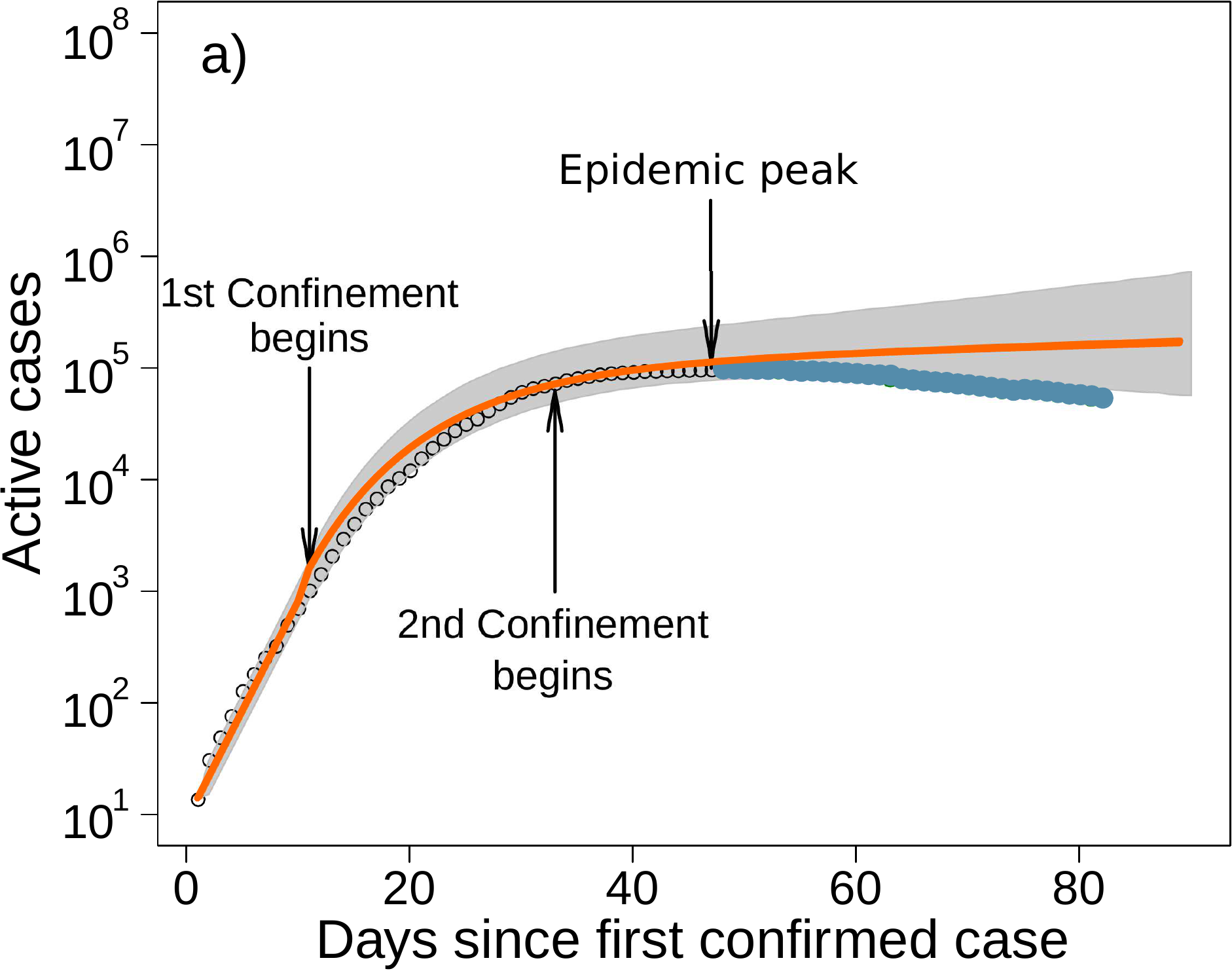}\\ 
		\includegraphics[width=0.45\textwidth,clip=]{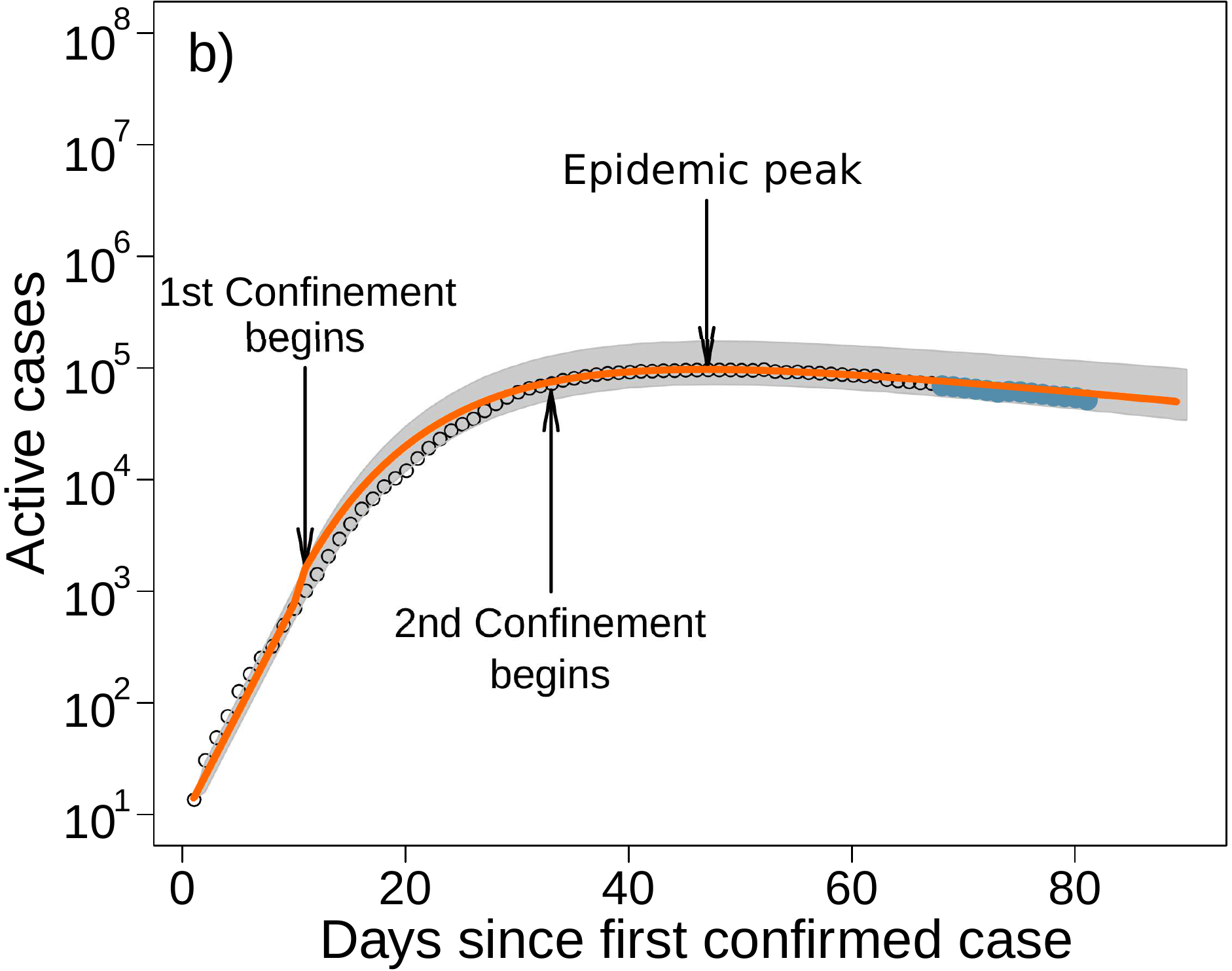}\\
		\includegraphics[width=0.4\textwidth,clip=]{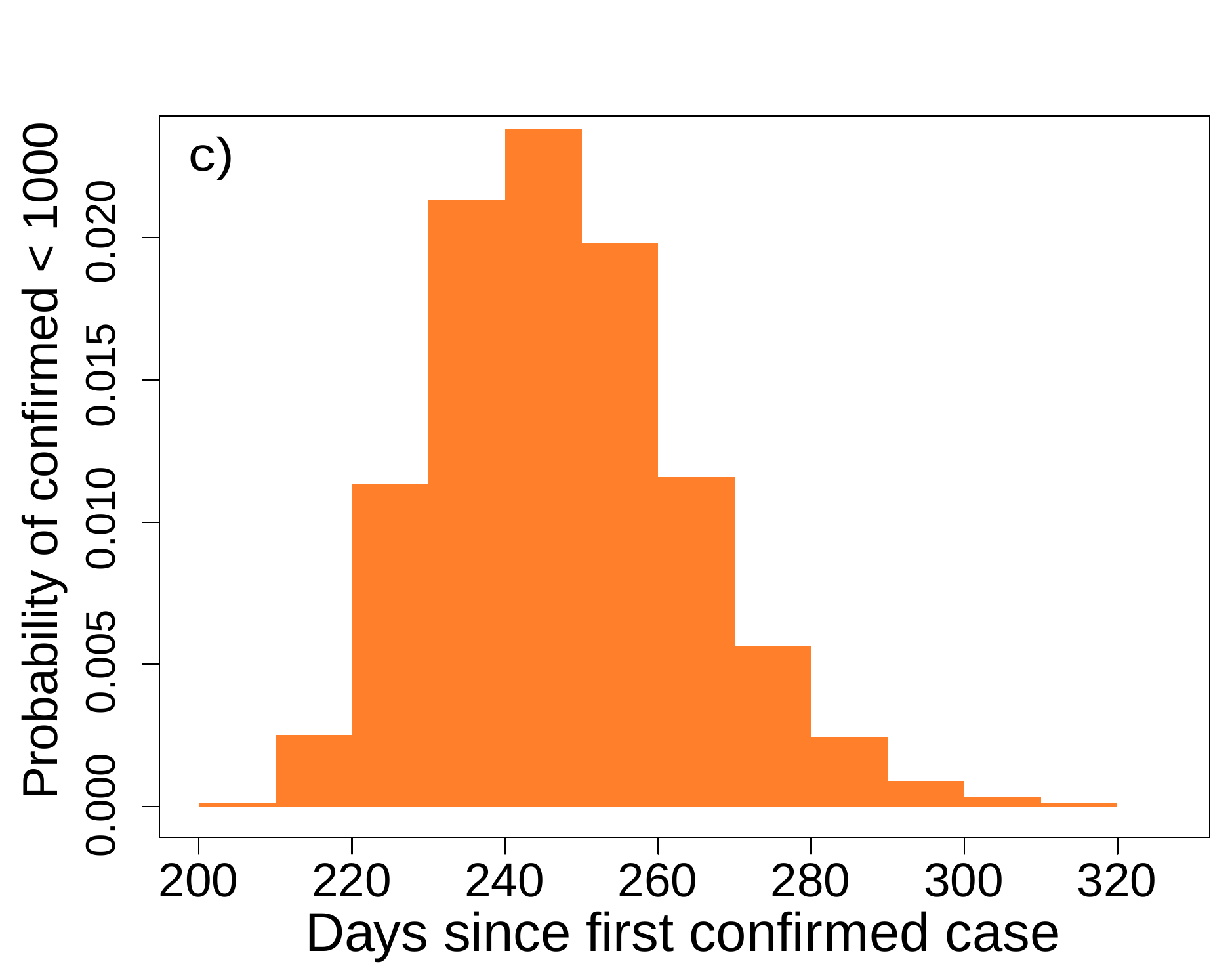}
    \caption{\small Fit to post-peak data for the daily number of active cases in Spain. (a) Fit to data up to April 18th (peak day). (b) Fit to data at three weeks post peak (May 9th). Open symbols represent fitted empirical data and blue dots correspond to actual measurements until May 17th. 
    (c) Distribution of times until the number of confirmed cases falls below 1000 for the first time. With about 2 cases per million inhabitants, this threshold can define the end of the epidemic. The distribution spans about three months centered around the end of October 2020.}
    \label{fig:postpeak}
\end{figure}

The peak of the epidemic in Spain was actually attained around April 18th. In the light of the broad distribution of times returned by fitting pre-peak data, and the forecast of only a 26\% probability of having a peak, the application of more stringent confinement measures on March 30th seems justified, and may have played an important role in the inhibition of COVID-19 propagation. Once the peak has been overcome, however, the uncertainty regarding the end of the epidemic remains high in the mid-term. Figure \ref{fig:postpeak} displays fits of data and forecasts just when the peak was attained and until May 9th using, as above, the analytical result in \eqref{eq:linearSQX}. The model is able to fit well past data and yields apparently narrower error intervals, as compared to pre-peak fits. 
However, data deviate from the most likely forecast about two weeks into the future. Unfortunately, the publication of data on the number of recovered cases was interrupted on May 17th, making it impossible to know the number of active cases, so we cannot extend the empirical series to the time of writing.
As of July 2020, Spain was experiencing a long plateau with a sustained average number of new daily cases well captured by the slow decrease in the $I(t)$ curve. However, even if the progression of the disease were well described by Figure \ref{fig:postpeak}(b), the distribution of times at which the epidemic would end is very broad.
Figure~\ref{fig:postpeak}(c) shows that there is an uncertainty of around three months in the time needed for the number of confirmed cases to drop below 1000---hardly an accurate prediction of the epidemic's end.

\subsection*{Better data are necessary, but not sufficient}

There are major difficulties in predicting the future of an epidemic with the present or similar models. First, incomplete or noisy data entail uncertain predictions, as can be seen in fits to all Spanish Autonomous regions, Fig.~S5. But, even if data were complete and precise, small variations in the parameters bring about growing uncertainties as time elapses. In order to illustrate this point, we have generated a synthetic set of observations through direct integration of the system described in Figure~\ref{fig:diagram}. By construction, uncertainties in the prediction of future trends derived through the Bayesian approach can only be ascribed to the dispersion of posterior distributions (shown in Fig.~S8).

\begin{figure}[!htp]
    \centering
    \includegraphics[width=.49\textwidth]{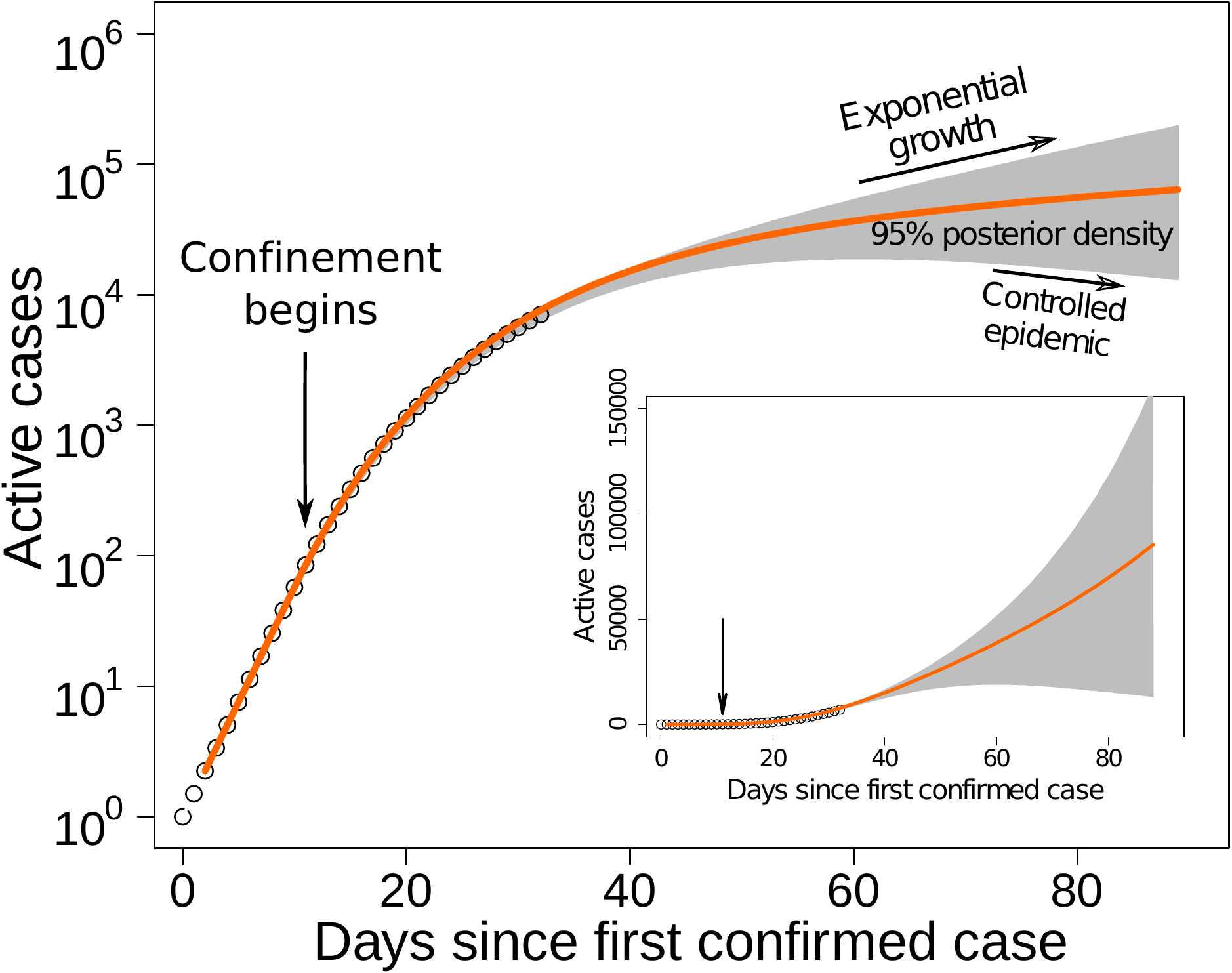}
    \caption{\small Data generated through direct simulation of the system described in Figure~\ref{fig:diagram} are used as input to determine posterior distributions for parameters through a Bayesian approach. Parameter values are $\beta=0.425$, $p= 0.007$, $q=0.062$ and $r+\mu= 0.021$, in the mitigation regime taken from the median of the posteriors in Fig.~S1 in SI (all measured in day$^{-1}$). Though the data set is complete and noiseless, consideration of only the growing phase of the epidemics implies a remarkable uncertainty in compatible trajectories. It is worth noting that, albeit those parameters would predict that the epidemic is not controlled, variability still leaves 3\% chance that it actually is. Inset: same data and curves with linear vertical scale.}
  \label{fig:active-mock}
\end{figure}

Figure~\ref{fig:active-mock} summarizes the inherent limitations of our model (and any other SIR-like model) to forecast any quantity beyond a limited temporal threshold. Prediction is severely affected by small differences in the fitting parameters, yielding a fan of compatible trajectories and limiting the reliability of the forecast. To complicate things further, the uncertainty of forecasts is not a monotonous function of the number of data points used in the model fits. Figures S9 and S10 illustrate the breadth of the 95\% posterior density as a function of the number of data points used, both for empirical and synthetic data. They show that more data does not necessarily entail less uncertainty and improved predictability.

\section*{Discussion}
\label{sec:discussion}

\subsection*{Confinement and turning points}

The implementation of confinement measures to control the expansion of highly transmissible pathogens affects the speed of infection propagation, as measured in the number of newly infected, recovered or deceased individuals. Confinement bends the progression curve downwards, but this bending, which can span a remarkable lapse of time, should not be interpreted as an unequivocal sign that propagation is to be inhibited. Rather, it might represent just a transient, cross-over regime to a new diverging, exponential phase, albeit with a different coefficient. In the simple SCIR model here discussed, these two regimes are clearly identified. The initial growth, before confinement starts, occurs at a rate $(R_0-1)(r+\mu)$, which depends on intrinsic properties of the pathogen-host interaction and on contacts between hosts. Sufficiently severe infections, with $R_0>1$, cause a pandemic if not controlled. The onset of confinement modifies the long-time trend of the infection by defining a new coefficient for the asymptotically dominating exponential, $(R_0 p(q+p)^{-1}-1)(r+\mu)$, which includes two important factors: the strength of confinement measures, $q$, and the lack of adherence of individuals to confinement, $p$. If $R_0p(q+p)^{-1}>1$ the growth slows down before a new asymptotic phase of exponential growth sets in. This phase corresponds to mitigation of infection propagation, but eventual extinction will occur through the usual SIR mechanism of exhaustion of susceptible individuals, while the deceleration observed in between is just a cross-over between two exponentially diverging regimes.
If the proportion of infected becomes high enough during the cross-over, our approximation is no longer valid and the SIR mechanism can kick in, making the second exponential regime unobservable in practice.

Inhibition of infection propagation is achieved only if $R_0p(q+p)^{-1}<1$, where a limited fraction of the population---which depends on the confinement strength and collective adherence---will get the disease. Though the model we have studied here is simple enough so as to allow us to derive exact results and to characterize the nature of the turning point (or cross-over), the qualitative scenario should be shared by any other compartmental models with growing and decreasing phases dominated by exponential behavior and able to implement the effects of confinement.

Phenomenological models do a good job in reproducing the temporal track of single outbreaks that result either from the unconstrained SIR mechanism of exhaustion of susceptibles or from a sustained inhibition of propagation down to extinction. However, models such as the generalized logistic \cite{wu:2020} or Gompertz \cite{levitt:2020} growth curves cannot capture some of the important qualitative behavior in the SCIR class. Those models cannot reproduce long plateaus such as those extending for months in Spain (see Fig. S11) or Italy. Phenomenological models do not have instability points such as a mitigation-inhibition threshold and, for that reason, they cannot embrace the possibility that propagation is slowed down but not halted (see Supplementary Information). The identification of such thresholds is essential to quantify the effectiveness of non-pharmaceutical interventions \cite{dehning:2020}.

\subsection*{Related SIR-like models}

There are several models in the literature conceptually analogous to the one described above. Obvious ones are SIR and SEIR (where the $R$ state is understood as ``removed'' individuals and groups both recovered and dead individuals). In SEIR models, the $E$ state permits to include the effect of a latent period where individuals are infected but asymptomatic. Depending on the disease it mimics, individuals can be infectious or not. In general, the consideration of the $E$ state brings about a delay in the completion of the disease, but does not entail qualitative changes. 

Some models consider the effect of quarantined individuals. SIQR and SIQS models have been introduced and studied early \cite{feng:1995,hethcote:2002}, but confinement was not considered there. More recent models aiming at including the effects of confinement and quarantine as applied in Italy and Spain to contain COVID-19---i.e., to class $S$---have generalized the classical SEIR model \cite{peng:2020,caminobeck:2020} in a way different from ours.
In one model \cite{peng:2020}, the possibility of individuals not committing themselves to the confinement (analogous to setting $p\rightarrow 0$ in our model) is discarded, while possible advances in treating the disease through a recovery rate that increases with time and a death rate that decreases with time are included. A variant of such model was used to draw very precise predictions on the course of epidemics that had to be subsequently revised in the light of new data \cite{lopez:2020}. In another case \cite{caminobeck:2020}, non-simultaneous in-going and out-going fluxes to the $S$ compartment are considered, thus standing for strict confinement followed by free relaxation.
A so-called SIRX model with irreversible quarantine has been shown to recover the systematic sub-exponential growth in the expansion phase \cite{maier:2020}. However, the quarantined class acts as an absorbing state and leads to an unrealistic feature of the model: at any strength, quarantine entails inhibition of infection propagation. This property is equivalent to the saturation implicit in phenomenological models, which cannot embrace mitigation of epidemic propagation without full inhibition (see SI, Section D). Finally a model quite similar to ours has been proposed, though no results on its dynamics are available as of yet \cite{MUNQU:2020}.

Still, other models increase the number of different states considered with the aim of becoming more realistic, especially motivated by specific observations of COVID-19. For example, the fact that only a fraction of the actually infected individuals is detected has been included in generalized mean-field models \cite{ivorra:2020}, or the different progress of the disease depending on the age group has been taken into account in models that consider stratified, age-structured populations with \cite{prem:2020} or without the consideration of physical location \cite{perez-garcia:2020}. Those models, by definition, have a significantly larger number of parameters. Again, the eventual aim of those models is to draw apparently soundly motivated and precise predictions on the time at which the pandemic will halt.

Compartmental models are appealing at least for two reasons: they are simple to formulate and offer a clear epidemic interpretation. But different models---as one can see by comparing the examples in this section---lead to different predictions. All models use either observational data on the progress of the epidemic or empirically evaluated parameters, or both. However, many of them lack a sensitivity analysis that propagates actual errors in data and parameters to their predictions. It is to be expected that, should they do so, most predictions might turn out to be compatible with different models (with different effective parameters) within their intervals of confidence. These models might return some reliable, probabilistic forecasts in the near future, but only if current conditions for propagation (e.g. confinement measures or the average collective habits of the population) remain unaltered \cite{wong:2020}. The reconstruction of the Wuhan epidemic, for instance, clearly illustrates that changes in contention measures turn reliable prediction nearly impossible \cite{hao:2020}.

\subsection*{Effective parameters and identifiability}

In principle, parameters characterizing the transitions between states in any SIR-like model are related to quantities amenable to empirical estimations. For instance, $\beta$ quantifies the transmissibility of the virus, $q$ should relate to the fraction of confined population, and vary with different non-pharmaceutical measures put in place; $p$ quantifies the adherence of population to confinement rules, and thus can be estimated through data on mobile phone location \cite{google:2020}; and so on. However, it is important to emphasize that a direct empirical estimation of the parameters need not be the right value to input the model with. Any simple SIR-like model (and even more so any phenomenological model) is by definition leaving aside a number of realistic features. When fits of actual data are attempted, those models yield at best trajectories close to the actual one, but the inferred parameters are necessarily effective. Including or not the $E$ state, confined individuals, age-structured populations or any other possible level of detail redefines the precise meaning and values of the corresponding rates. Actually, an SIR model has been shown to reproduce better data from COVID-19 spread in Wuhan than an SEIR model \cite{roda:2020}, as well as to be ``unreasonably effective'' in describing the outbreaks in different countries affected by the pandemic \cite{carletti:2020}. Moreover, the choice of which states and how many of them to include in the model is a subjective matter that changes dramatically the quality of the fit and, more importantly, the interpretability and identifiability of the parameters~\cite{beauchemin2017duration}.

It could be argued that more realistic models are those with a larger number of states and parameters. But, at the same time, in those models it is more difficult---often impossible---to assign a unique meaningful value to their parameters. If, rather than using the empirical estimates themselves, we attempt a non-informed fitting to the data, we normally end up in a problem of identifiability. That is, there are many different parameter sets (often a continuum of them) that fit the data, a problem that has been specifically analysed for SIR-like models applied to COVID-19 \cite{massonis:2020}. Though the use of empirical values from independent studies might anchor the values of a subset of parameters, the remaining degrees of freedom (parameters with values not fixed \emph{a priori}) are often sufficient to reproduce data. In this sense, multi-parametric models might lead to over-fitting, loosing at the same time explanatory power. 

\subsection*{Probabilistic forecasting}

Deterministic epidemiological models convey a false impression of uniqueness of trajectories. It is broadly believed that a model able to reproduce empirical data well should be equally good at predicting future outcomes. This causal fallacy too often prevents a careful evaluation, through sensitivity analyses, of the effects of small variations in the parameters in forecasting. When such analyses are performed, results similar to ours should be obtained. 

More detailed models using SIR-like descriptions for metapopulations and adding mobility usually incorporate uncertainty in their predictions. In this way, ranges for different quantities have been derived, such as for the basic reproductive number $R_0$ and its temporal response to confinement measures \cite{arenas:2020b}, for the number of undocumented infected individuals \cite{li:2020_Sci} or estimations of the effect of physical distancing measures in the median number of infections \cite{prem:2020}. 

At the other end of the spectrum one finds statistical approaches that, in the absence of an underlying mechanistic model, fully rely on past data to predict the near future. Numerical approaches of this kind are intrinsically probabilistic and only yield likelihoods of different scenarios, with intervals of confidence that grow extremely fast as time elapses. A notable example of such approaches is the document elaborated by researchers at Imperial College \cite{flaxman2020estimating}, based on Bayesian estimations only informed by Europe-wide data of the COVID-19 pandemic. The results are compatible with multiple scenarios in most countries in the mid-term, since the huge intervals of confidence inherent to their approach limit predictability to the near future. At the same time, that document likely yields the most trustworthy (probabilistic) predictions using pre-peak data.
The use of recent techniques that incorporate the possibility of obtaining closed mathematical expressions to Bayesian approaches \cite{guimera:2020} might bridge those techniques and the identification of underlying mechanistic models. 

\section*{Conclusions}

SIR-like models are unable to predict with certainty; at most, they can inform on the different likelihood of a variety of trajectories conditional on specific measures and parameters. Uncertainties in the values of the latter prevent a unique interpretation of the data at the transient. Near the threshold separating mitigation from inhibition, the same set of observations might be compatible with either future outcome. If the aim of control protocols is to minimize the total number of infected individuals and the duration of the confinement period, it seems advisable that the strongest possible measures are applied as early as possible. It has been documented that non-pharmaceutical interventions during the 1918 Flu Pandemic in the U.S. lowered mortality and mitigated adverse economic consequences \cite{correia:2020}. Deferral of such application is not justified on the basis of a slow-down of infection propagation. 

Lorenz closed his 1972 talk \cite{lorenz:1972} by stating:
\begin{quote}
``[Errors in weather forecasting] arise mainly from our failure to observe even the coarser structure with near completeness, our somewhat incomplete knowledge of the governing physical principles, and the inevitable approximations which must be introduced in formulating these principles as procedures which the human brain or the computer can carry out. These shortcomings cannot be entirely eliminated, but they can be greatly reduced by an expanded observing system and intensive research. It is to the ultimate purpose of making not exact forecasts, but the best forecasts which the atmosphere is willing to have us make that the Global Atmospheric Research Program is dedicated.''
\end{quote}
Could COVID-19 trigger a Global Epidemic Research Program, intensive investigation in the topic and an expanded observing system producing accurate and publicly available data? Only through such a Program would it be possible to obtain the best forecasts that epidemic models might yield. 

\section*{Materials and Methods}

\subsection*{Data} We have used two slightly different datasets for studies pre- and post-peak. The first one was downloaded in real time as the epidemic progressed in Spain and contains information on number of new infected, dead and recovered cases as it was available by the 1st of April, 2020. The second dataset, used in the post-peak analysis, contains data updated in retrospect by the Spanish Ministry of Health taking into account modifications in the official reporting introduced since April 2020. All data and codes used in this work, together with explanations on how to retrieve our fits, are publicly available at \url{https://github.com/mariocastro73/predictability}.

\subsection*{Bayesian fit} We have fitted a parametric Bayesian model with the variables
\begin{equation}
\begin{split}
& \log I(t) \sim {\mathcal N} \big( \log I(t_0)+(\beta-\lambda)(t-t_0),\sigma_I \big), \ t_0<t\leq t_2, \\[-1mm]
& \log I(t) \sim {\mathcal N} \big(\log I(t_2)+ [R^*_0(t-t_2)-1]\lambda(t-t_2),\sigma_I \big), \ t_2<t, \\[-1mm]
& \log X(t) \sim {\mathcal N} \big( \log(r+\mu)+\log I(t),\sigma_X \big), \ t_1<t, 
\end{split}
\label{eq:Bayesian1}
\end{equation}
where $X(t)=\Delta R(t)+\Delta D(t)$ stands for the change in the number of recovered plus dead cases daily reported. Similarly, we choose the following priors
\begin{align}
& \beta \sim {\mathcal U}(0,1), && 1/\sigma_I^2 \sim \Gamma(0.01,0.01), \nonumber \\
& r+\mu \sim {\mathcal U}(0,1), && 1/\sigma_D^2 \sim \Gamma(0.01,0.01), \label{eq:Bayesian2} \\
& p \sim {\mathcal U}(0,5), && 1/\sigma_X^2 \sim \Gamma(0.01,0.01), \nonumber \\
& q\sim {\mathcal U}(0,5), && \nonumber
\end{align}
where $\sim$ stands for \textit{distributed as}, and $\mathcal N$, $\mathcal U$ and $\Gamma$ stand for Normal, Uniform and Gamma distributions. For each Spanish Autonomous region, $t_0$ and $t_1$ stand for the days (since February 28th) where the first Infected and Recovered+Death cases were reported, respectively.

Note that we distinguish between the epidemic before ($t\leq t_2$) and after ($t>t_2$) any confinement measure was applied. Unlike other implementations of epidemic models~\cite{flaxman2020estimating} our model in \eqref{eq:Bayesian1} aims to fit directly the solution of the deterministic model in \eqref{eq:linearSQX}.

The priors for $\beta$ and $r+\mu$ in \eqref{eq:Bayesian2} are informative priors derived from the fact that in every country where cases of coronavirus are detected, the doubling period at the very early stages of the epidemic was never smaller than two days. Hence we have taken for these parameters a prior $\mathcal{U}(0,1)$. For the other parameters we assume a non-informative prior $\mathcal{U}(0,5)$ (we assume that changes faster than $1/5$ day are meaningless in any compartmental model). The results are consistent with this assumption.

To profit from the largest amount of available data, we fit simultaneously the number of active cases ($I$) and the new number of cases in the interval $\Delta t=1$ day of recovered ($R$) and dead ($D$) cases in logarithmic scale---a linear scale would give a biased larger weight to later times.

We have implemented the Bayesian model in R using the rjags wrapper of the JAGS library~\cite{plummer2003jags}. The full code is provided in Sec.~B 
of the SI and can be downloaded from \url{https://github.com/mariocastro73/predictability}.

\section*{Acknowledgments}
{The authors are indebted to Dami\'an H. Zanette for his critical reading of a previous version of this work, and to Jacobo Aguirre, Yuxuan Cheng, Robert Endres, Javier Mart\'{\i}n-Buld\'u, David J. J\"org, and Anxo S\'anchez for their useful comments. This research has been funded by the Spanish Ministerio de Ciencia, Innovaci\'on y Universi\-da\-des-FEDER funds of the European Union support, under projects FIS2016-78883-C2-2-P and PID2019-106339GB-I00 (M.C.), BASIC (PGC2018-098186-B-I00, J.A.C.), MiMevo (FIS2017-89773-P, S.M.),  PerIODIC (FIS2016-78313-P, S.A.) and BADS (PID2019-109320GB-100/AEI/10.13039/501100011033, S.A.). The Spanish MICINN has also funded the ``Severo Ochoa'' Centers of Excellence to CNB, SEV 2017-0712, and the special grant PIE 2020-20E079 (CNB, S.M. and S.A.) entitled ``Development of protection strategies against SARS-CoV-2''.}



\end{document}